\documentclass[graybox]{svmult}

\usepackage{mathptmx}       
\usepackage{helvet}         
\usepackage{courier}        
\usepackage{type1cm}        
\usepackage{makeidx}         
\usepackage{graphicx}        
\usepackage{multicol}        
\usepackage[bottom]{footmisc}

\makeindex             

\begin{document}

\title*{Preliminary results of Ion Backflow study for a single GEM detector}
\author{Deb Sankar Bhattacharya$^a$, P K Sahu$^a$, Sagarika Swain$^a$, Sanjib K Sahu$^a$}
\institute{Deb Sankar Bhattacharya, \email{dsb.physics@gmail.com}
\\(a) Institute of Physics, Sachivalaya Marg, Bhubaneswar, Odisha 751005.
}
%
%
\maketitle

\abstract{Gaseous detectors are used in both low energy and high energy physics experiments. The present day gaseous detectors, i.e., the Micro-Pattern gaseous detectors (MPGD) are more efficient and fast. Gas Electron multiplier (GEM) is quite well know among the MPGD members. The MPGDs are also being used in other applications like tomography/imaging, moreover, recently, hybridization of two different kinds of MPGD is another emerging subject of R\&D. Ion feedback is an intrinsic drawback of the gaseous detectors. However, it is not a big issue where the event rate is not very high, or the drift volume is not too large. Here, we are showing a simple experimental technique to find the ion feedback of a single GEM foil. This can address the experiments/applications where a single GEM foil is employed.} 

\section{Introduction}
Among many other Micro-Pattern Gaseous Detectors (MPGD) \cite{ref2-MPGD}, the Gas Electron Multiplier (GEM) \cite{ref2-GEM-Sauli} is
specially remarkable for very good position resolution, energy resolution, stable high gain and low ion feedback. For its excellent performance, the GEMs are being adopted in many HEP experiments such as in ALICE, CMS, CBM etc. ALICE \cite{ALICE-TDR} has reported to upgrade the gas-amplification technology of their Time Projection Chamber (TPC) \cite{ref2-TPC-Nygren-1975, ref2-TPC-Hilke-1975} from Multi-Wire Proportional chambers (MWPC) \cite{ref2-MWPC-Charpak} to GEM to cope with the high-multiplicity
environment after Long Shutdown 2. GEM has also been conceived as an efficient candidate for the ILD-TPC for the upcoming ILC \cite{ILC}. Ion feed back is a general issue of all the MPGDs. According to the physics goals and the detector structures, IBF in GEM is being studied by different experimental groups.  
 
In this report, we present the procedure of assembling a single GEM, preparation of the experimental setup and the measurement of ion backflow fraction for it. This gives the insight of how the ion backflow of a single GEM foil behaves in different electric field and voltage configurations. 


\section{Gas Electron Multiplier}
\label{sec:1}
A GEM comprises a polymer foil, with copper coating on
both the sides. The foil is patterned with a matrix of
identical holes, typically 50-100 per mm$^2$ (Fig. 1a). Piercing on the foil is done using chemical etching and
photolithography technique. The shape and the pitch of the holes may vary.
A typical and widely used shape is double-conical shape (Fig. 1b). However, in our experiment, we are using a single mask GEM foil. The outer and inner hole diameters are respectively 70 $\mu$m and 50 $\mu$m. The hole pitch of the foil is 140 $\mu$m.
In a closed volume, when the radiation ionizes the gas,
by applying suitable potential difference between the two
copper sides of a GEM foil, the primary electrons are collected,
multiplied and then driven towards the signal collection plane.

\begin{figure}[b]
\includegraphics[scale=.40]{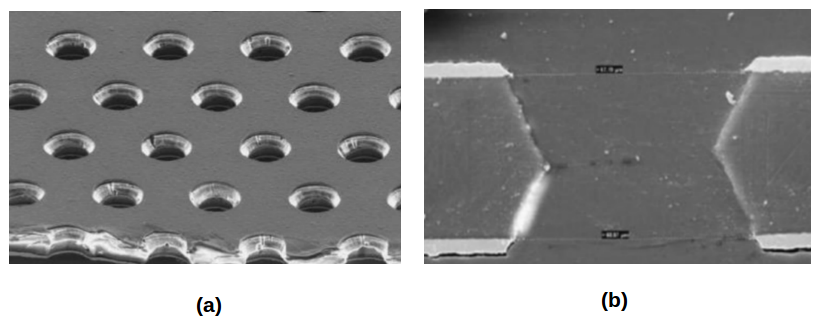}
%
%
\caption{(a) Microscopic picture of a GEM foil. The holes can be seen. (b) Cross section of a GEM hole. The double-conical shape can be seen. There are two layers of coppers on the top and on the bottom. There is the polymer layer at the middle. \cite{Sauli Paper 2016}}
\label{fig:1}       
\end{figure}

\section{Investigation on a single GEM foil}   
\label{sec:3}
With a motivation to comprehend the intrinsic ion suppression property, we started our investigation with a single GEM foil as a first step at the Institute of Physics, Bhubaneswar (IOPB). We stretched a GEM foil and assembled a chamber of one GEM. The preparation of the setup, experimental procedure, and the results are discussed here.

\subsection{Assembling the single GEM}   
\label{sec:3.1} 
The GEM foils and the other necessary components were procured from CERN. For thermal stretching of the GEM foil, we devised two 30 cm $\times$ 30 cm acrylic frames (Fig. 2a) at the IOPB-workshop. The two halves of the frame are attached with the help of 20 equispaced bolts to ensure uniform tension on the GEM foil. After the GEM foil is placed between the frames and the screws are uniformly tightened, the system is heated in a controlled way. A Halogen lamp is used as the heat source (Fig. 2b). A thermometer is kept nearby to monitor heat. A glued-frame which can grip the GEM foil is then placed on the stretched foil when the temperature is between 42$^{\circ}$ C to 44$^{\circ}$ C. Another glued-frame is also placed instantaneously on the other side of the GEM foil. That is how, the GEM foil is stretched and fixed between two glued-frames, each 0.5 mm thick. Following the same procedure, the drift plane is also stretched. The drift plane is basically 5 $\mu$m copper plated Kapton foil of 50 $\mu$m. 

The anode plate has 4 holes at the corners to insert 4 nylon studs. A stack of 3 nylon spacers, each 0.5 mm thick are placed around each of the 4 studs. Then the GEM-frame is gently placed on top of the spacers, thus creating an induction gap of 2.0 mm (the glued-frame is 0.5 mm thick). After that, the cathode plane (or the drift plane) is placed 3.5 mm above the GEM foil with the help of 5 spacers. When all the setup is ready with the desired induction gap and drift gap, a solid frame of G10 material is placed around it. The G10 frame has 2 O-rings on top and bottom sides of it. It is firmly mounted on the anode plane with the help of 28 metal screws. On the top side of the G10 frame, a 100 $\mu$m thin Kapton window is placed. Then all the metal screws are gently tightened. Finally, to prevent any leak, araldite glue is applied on the anode sides of the plastic studs. 

\begin{figure}[b]
\includegraphics[scale=.42]{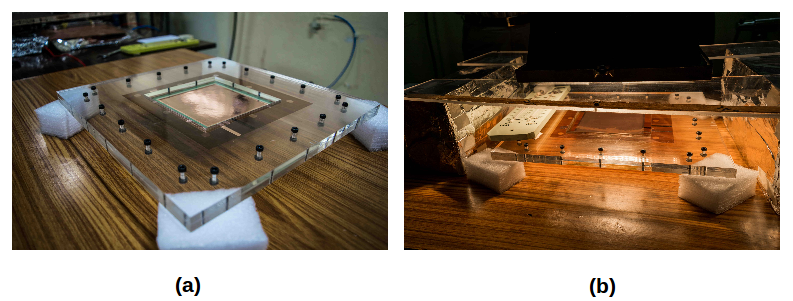}
%
%
\caption{(a) The The Acrylic Jig. (b) Thermal stretching of a GEM foil. A Halogen lamp of 1000 W as heat source and a thermometer.}
\label{fig:3}       
\end{figure}

\subsection{Experimental setup}   
\label{sec:3.2}

The ion backflow fraction is defined as the ratio of the number of ions drifting back to the cathode to the total number of ions produced during avalanche. The avalanche occurs inside the GEM holes and the electron are partially collected at the bottom part of the GEM, and the rest at the anode. 
Now, the total number of ions present inside the GEM holes must be equal to the sum of the numbers of the electrons collected at the anode and at the bottom part of the GEM. 
Therefore, the ion backflow fraction can be written as,
\begin{equation}
IBF = \frac{I_C}{I_B+I_A} 
\end{equation}
where, $I_C$ is the cathode current (or drift current), $I_B$ is current from the bottom part of the GEM, $I_A$ is anode current. 

To measure the currents from different electrodes of the setup, we used Keithley (6485) pico-ammeter. For the ionizing radiation, we used Cs$^{137}$ isotope which emits 0.66 MeV gamma photon. It is quite obvious that to measure IBF, one needs to measure currents from different components, even when they are kept at certain potential. Rather measuring current from a floating component, we applied a different procedure. 

\begin{figure}[b]
\includegraphics[scale=.42]{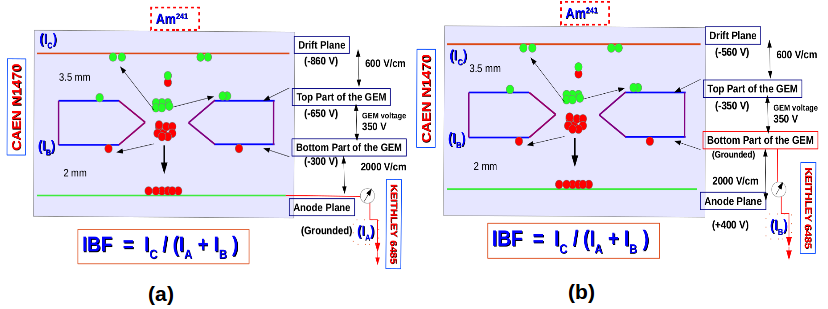}
%
%
\caption{In these two configurations, the drift field is 600 V/cm, GEM voltage is 350 V, and induction field is 2000 V/cm. Only the way of applying potentials have been varied. (a) The configuration for measuring $I_A$ (b) The configuration for measuring $I_B$.}
\label{fig:4}       
\end{figure}


\begin{figure}[b]
\includegraphics[scale=.41]{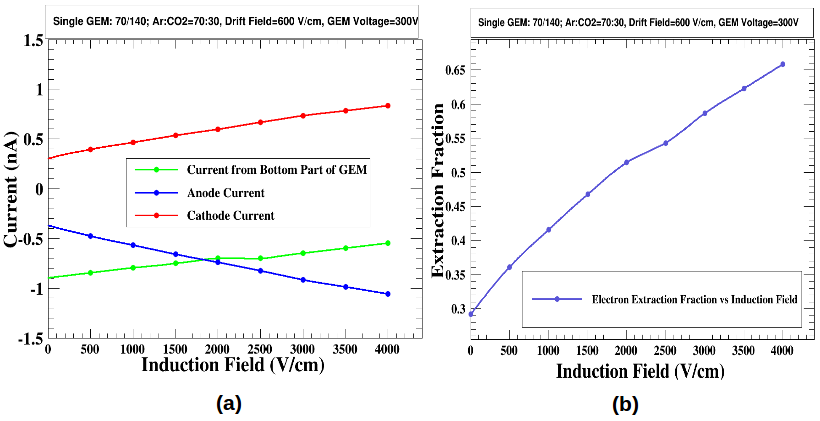}
%
%
\caption{(a) Change of cathode current, anode current and current from the bottom part of the GEM is varying with induction field. (b) Change of extraction fraction (equation 2) with induction field.}
\label{fig:5}       
\end{figure}


\begin{figure}
\centering 
\includegraphics[scale=.19]{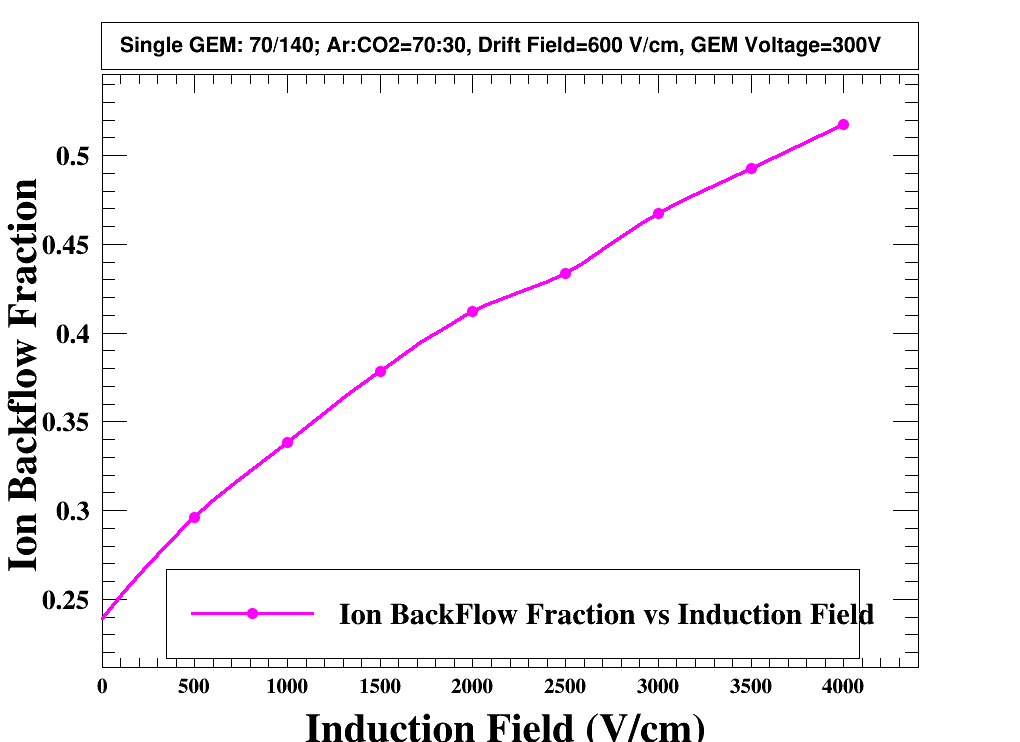}
%
%
\caption{Change of Ion Backflow Fraction with induction field. Ion backflow increases with induction field.}
\label{fig:6}       

\end{figure}


\begin{figure}[b]
\includegraphics[scale=.41]{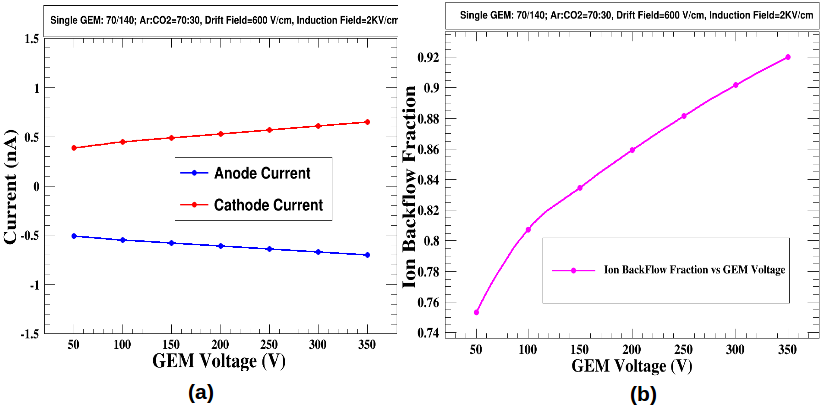}
%
%
\caption{(a) Change of cathode current, anode current with varying GEM voltage. (b) Change of Ion Backflow Fraction with GEM voltage.}
\label{fig:7}       
\end{figure}


\begin{figure}[b]
\includegraphics[scale=.41]{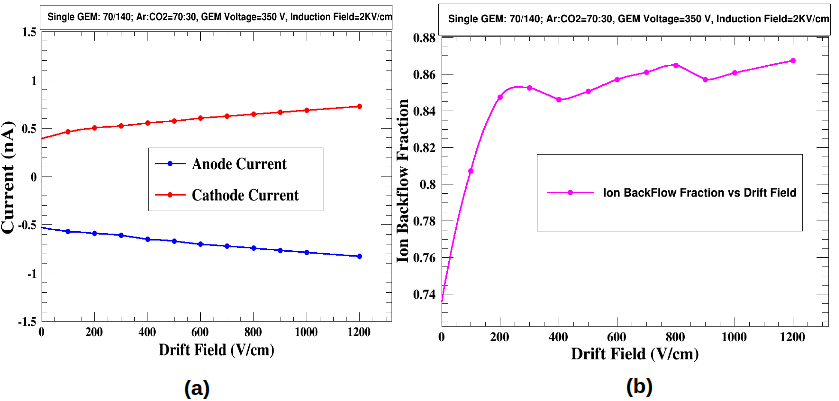}
%
%
\caption{(a) Change of cathode current, anode current with varying drift field. (b) Change of Ion Backflow Fraction with drift field}
\label{fig:8}       
\end{figure}


Generally, in normal mode of operation, the anode plane of a GEM is grounded (Fig. 3a). The bottom part of the GEM, the top part of the GEM and the cathode are kept at negative potentials, according to the desired drift field, GEM voltage, and induction field. Therefore, by simply connecting the ammeter to the anode plane will give $I_A$. Now, to measure current from the bottom part of the GEM, $I_B$, we biased the anode plane to a positive potential and grounded the bottom part of the GEM. The top part of the GEM and the cathode is biased accordingly so as to keep the drift field and the GEM voltage and induction field exactly the same as before (Fig. 3b). Likewise, when measuring current from the cathode, $I_C$, the cathode is grounded and all the other electrodes are biased with positive potential in such a way that the electric fields remain the same as before.  

In Fig. 3, two different configurations are shown. These two configurations of potential are adopted to measure $I_A$ and $I_B$. It should be noted that the field configurations for them are exactly the same. With a similar configuration, current from the cathode, $I_C$ can also be measured. During all the current measurements, the temperature was maintained at around 20$^{\circ}$ C and the pressure was around 1001 mBar. Also, the current from the ammeter is noted for 2 mins to ensure that the value is stable.  

\subsection{Results}
\label{sec:3.3}
Following the described method, we measured currents from anode, bottom part of GEM and from cathode. A scan of the induction field is done and variation of all the currents are noted. The variation is presented in Fig. 4a. The sum of the currents from the bottom part of the GEM and from the anode gives the total number of electron produced, while current from anode only is responsible for the signal induction at the readout plane. Interestingly, it may be noted here that the magnitude of $I_B$ is decreasing, while the magnitude of $I_A$ is increasing with the induction field (both $I_B$ and $I_A$ are in negative scale in Fig. 4a). The result can be explained easily: after the avalanche inside the GEM holes the electrons are attracted towards the anode by the induction field, however, not all of them manage to reach the anode. Immediately after exiting the GEM holes, a fraction of the total number of electrons terminate to the bottom part of the GEM. This fraction is measured as $I_B$ and it also decides the electron extraction efficiency from the GEM holes. As the induction field increases, the collection of electrons on the bottom part of the GEM becomes more and more smaller, since they are more driven towards the anode. Hence the anode current increases. On the other hand, the cathode current slowly increases with the induction field as the ion extraction from induction gap also increases with induction filed. If the sum of $I_B$ and $I_A$ gives the total number of electrons present after avalanche, an extraction fraction (extraction of electrons from GEM holes) can be defined as follows:
\begin{equation}
\epsilon_{extraction} = \frac{I_A}{I_B+I_A} 
\end{equation}
In Fig. 4b, the change of extraction fraction with induction field is shown. At an induction field of 2000 v/cm, the extraction fraction is nearly 50$\%$. Finally, in Fig. 5, the change of ion backflow fraction (IBF) with induction field is shown. Ion backflow is increasing with induction filed. The increase is guided by the changes of $I_C$, $I_B$, and $I_A$ which is presented in Fig. 4. 

Now, the voltage across the two copper sides of the GEM foil is varied and the currents from anode and cathode are measured. In Fig. 6a, the changes of anode current and cathode current are shown. Unfortunately the current from the bottom side of the GEM could not be measured in this case for a technical difficulty. It can be seen in Fig. 6a, that the magnitude of both the cathode and the anode currents are increasing. This is because, with increasing GEM voltage, the gain of the detector increases which means the total numbers of electrons and ions increase. In Fig. 6b, the change of IBF with GEM voltage is shown. In this IBF estimation, only the ratio of $I_C$ to $I_A$ is taken, $I_B$ is not considered. This only implies an assumption of 100$\%$ electron extraction from the GEM holes. If $I_B$ were considered, the IBF would be small than what is presented here.

After that, the drift field of the setup is varied and all the currents are measured. In Fig. 7a, changes of $I_A$ and $I_C$ are shown. With increasing drift field, the magnitudes of both the anode and cathode are increasing. The reason is, as the drift field increases, the collection of primary electrons in the GEM holes increases. It is obvious that more the electrons enter the GEM holes, the more will be the numbers of electron-ion pairs after avalanche. With the same assumption of 100$\%$ extraction efficiency, the change of IBF with drift field is shown in Fig. 7b. Here, IBF increases with increasing drift field and then trends to saturate. The saturation behavior of IBF can be seen as saturation in electron collection in the GEM holes with increasing drift field.   

Finally, with a Fe$^{55}$ source, which emits X-ray of 5.9 keV, we have measured the gain of the GEM. We have found that for 4 kv/cm induction field, 350 v GEM voltage and 800 v/cm drift field, the gain of the detector is around 1500 in the same gas mixture (Fig. 8). We could go upto a gain of 3500.   

\begin{figure}[b]
\includegraphics[scale=.30]{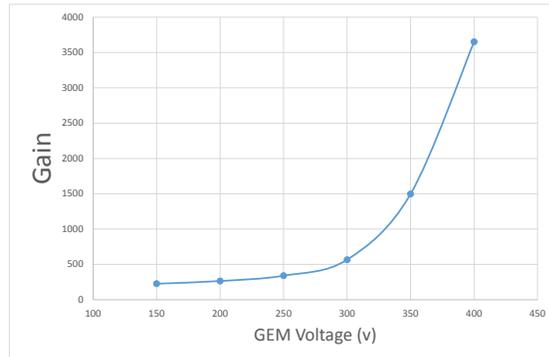}
%
%
\caption{The Gain of a single GEM foil is plotted for different GEM voltages. The drift field and the induction field are 800 v/cm and 4 kv/cm respectively.}
\label{fig:9}       
\end{figure}

 
\section{Conclusion}
Measurement of ion feedback of a single GEM foil is interesting for the experiments or applications which employs only a single GEM.
We assembled a chamber with a single GEM foil at the Institute of Physics with a motivation to estimate its ion backflow fraction. The experimental setup and the procedure for measuring ion backflow is discussed here in detail. Three main parameters: induction field, GEM voltage and drift field are individually varied (while keeping the other constant) and IBF is measured. The behavior of IBF while changing these parameters are explained. In some cases, while calculating IBF, we assumed an ideal situation of $100 \%$ electron extraction from the GEM holes.

\section{Acknowledgment}
We thank the members of IOP workshop for their invaluable cooperation.

%
%
%

\end{document}